\documentclass[reprint,amsmath,amssymb,prper,showkeys,floatfix]{revtex4-2} 
\usepackage[utf8]{inputenc}
\usepackage[ngerman, USenglish]{babel}

\usepackage{tikz,pgf,pgfplots}
\usepackage{tikz-3dplot,xcolor}
\usetikzlibrary{shapes.geometric,patterns,calc,matrix,arrows.meta}
\pgfplotsset{compat=1.18}
\definecolor{PRPink}{RGB}{157,91,179}

\usepackage{graphicx}
\usepackage{dcolumn}
\usepackage{bm}
\usepackage{docs}%
\usepackage{bm}%
\usepackage{multirow}
\usepackage{tabularx}
\usepackage{booktabs}
\usepackage{float}

\usepackage[normalem]{ulem}
\usepackage{dcolumn}
\usepackage{enumitem}
\usepackage[colorlinks=true,linkcolor=blue,citecolor=blue,urlcolor=blue]{hyperref}

\begin{document}
\title{A cross-age study on secondary school students' views of stars} 
\author{Philipp Bitzenbauer}
\email{philipp.bitzenbauer@fau.de}
\affiliation{Friedrich-Alexander-Universität Erlangen-Nürnberg, Professur für Didaktik der Physik, Staudtstr.~7, 91058 Erlangen, Germany}
\author{Sarah Navarrete}
\affiliation{Friedrich-Alexander-Universität Erlangen-Nürnberg, Professur für Didaktik der Physik, Staudtstr.~7, 91058 Erlangen, Germany}
\author{Fabian Hennig}
\affiliation{Universität Leipzig, Institut für Didaktik der Physik, Prager Str.~36, 04317 Leipzig, Germany}
\author{Malte S. Ubben}
\affiliation{Technische Universität Braunschweig, Institut für Fachdidaktik der Naturwissenschaften, Bienroder Weg 82, 38106 Braunschweig, Germany}
\author{Joaquin M. Veith}
\affiliation{Stiftungsuniversität Hildesheim, Institut für Mathematik und Angewandte Informatik, Samelsonplatz 1, 31141 Hildesheim, Germany}

\date{\today}
\begin{abstract}
Research in astronomy education has uncovered that many learners possess limited and fragmented understanding of stars. The corresponding misconceptions manifest in various areas such as star formation, size, the relationship between stars and planets, and their position in space and have been shown to persist across different age groups and educational settings, highlighting the need for further investigation. This paper presents the findings of an empirical study that examines secondary students' views of stars and their evolution throughout their secondary school careers. Therefore, we designed and evaluated an instrument for assessing students' views of stars in five domains (stars and the solar system, formation and evolution of stars, properties of stars, (sub-)stellar objects, and spectral aspects). The instrument creation process involved several steps, including literature-based item development, an expert survey with faculty members, and a quantitative pilot study with a sample of $N = 390$ secondary school and college students. This process led to a final version of the instrument that exhibits good psychometric properties. We used this new instrument in a cross-age study to investigate the alignment of secondary students' ideas about stars with scientific views across different stages of secondary education. The sample of this main study comprised a total of $N = 366$ learners, including 148 lower, 151 middle and 67 upper secondary school students. Our study findings reveal a progressive development of students’ perspectives on star-related topics throughout their school career: Using ANOVAs and conducting pairwise post-hoc comparisons, we observed a statistically significant increase in the proportion of responses aligning with scientific views across all aspects of stars examined in this study, as students progressed from lower secondary to upper secondary levels. We further report on widely held views of stars among our study participants that oppose the scientific views, and discuss the implications of our findings for both educational research and practice.

\end{abstract}
\maketitle

\section{Introduction} 
\label{sec:intro}
Learning about astronomical objects poses a challenge for individuals due to the limited, skewed, or lack of direct experience with these objects \cite{rajpaul2018cross}. Currently, research on mental models in astronomy education is mostly focused on topics related to celestial bodies such as the Sun, Moon, and Earth: For instance, there exists a large body of research on the development of students' mental models of Earth \cite{nussbaum1979children,kose2022gifted}, the origin of seasons \cite{baxter1989children}, the day-and-night cycle \cite{chiras2008day} or lunar phases \cite{stahly1999third}. In contrast, research on more elaborate and modern astronomy concepts is still in its infancy \cite{ubben2022holes,salimpour2023universe}, although astrophysics research findings are increasingly making their way into the spotlight of news broadcasts and newspaper articles due to recent advancements in the field (e.g., see \cite{tran2018effectiveness}), as evidenced by the multiple Nobel Prizes awarded for astrophysical research in recent years (e.g., 2002 \cite{Nobel2002}, 2006 \cite{Nobel2006}, 2011 \cite{Nobel2011}, 2015 \cite{Nobel2015}, 2017 \cite{Nobel2017}, 2019 \cite{Nobel2019} and 2020 \cite{Nobel2020}). However, there is no need to rely solely on the number of Nobel Prizes to gather unanimous support for the assertion that modern astrophysics is an exceptionally captivating realm. Physics education research has found that space and astronomy topics spark significant interest among both boys and girls \cite{holstermann2007interesse}, and this comes as no surprise as argued in Ref.~\cite{spatschek2003astrophysik}: From an early age, students are consciously confronted with astronomical questions through the inevitable act of observing the sky \cite{Meyn_2008}. Furthermore, their curiosity about the origins of humanity leads them to seek scientific explanations, highlighting the enduring allure and importance of astronomy and astrophysics in the eyes of students and non-specialists alike. Consequently, astronomy concepts are on the rise in K-12 physics education \cite{hennig2023big,Salimpour-2021}. Thereby, the topic of stars and their evolution holds significant potential for physics education from various perspectives: Stars are objects whose structure and life can be described through the interaction of different sub-disciplines of physics (for an overview, see \cite{bromm2013formation}). Through the application of fundamental physics principles, a comprehensive understanding can be gained of the formation and evolution of stars as well as their transition into compact objects such as white dwarfs, neutron stars, or black holes, which occurs when nuclear processes converting lighter elements (e.g., hydrogen and helium) into heavier elements (including carbon, oxygen, and elements across the periodic table up to uranium) cease. The example of stars offers a compelling demonstration of the interconnectedness between observation (such as spectral analysis) and theoretical description (for an overview, see \cite{hubeny2007stellar}). Hence, stars are not only a fundamental aspect of astrophysics but may also serve as a central theme in astronomy education: They provide an excellent opportunity for students to grasp core aspects of stellar evolutionary processes. Additionally, the theory of star formation remains incomplete, with unresolved questions, offering a platform for learning about the Nature of Science. For example, the mechanisms leading to the collapse of interstellar clouds, and hence to star formation, are still not fully understood \cite{fisher2014rarity}.

In this paper, we report on a cross-age study investigating secondary students' views of stars and their evolution throughout secondary school careers. Building on previous research on student learning of astronomy concepts in general, and stars in particular (see Section~\ref{sec:RB}), we formulate our research questions in Section~\ref{RQs}, and describe the design and evaluation of the research tool used in this study in Section~\ref{sec:design}. We present the findings of our study in Section~\ref{sec:results} and discuss these results against the backdrop of prior research in Section~\ref{sec:discussion}. Finally, we provide recommendations for both, future research and practice in astronomy education at the secondary school level (see Section~\ref{sec:conclusion}).

\section{Research Background}\label{sec:RB}

\subsection{The big ideas of astronomy education research}

\subsubsection{Topics covered and target groups addressed}
Lelliott and Rollnick \cite{lelliott2010big} conducted a comprehensive review of peer-reviewed astronomy education studies published between 1974 and 2008. Their review highlighted that the majority of these studies focused on teaching and learning of five big ideas, ``all involving the Earth in relation to its satellite and the Sun'' \cite{lelliott2010big} (p. 1777): (1) Earth \cite{nussbaum1979children,kose2022gifted,agan2004learning,sneider1998unraveling,jones1987children,vosniadou1992mental,nussbaum1976assessment,nussbaum1979children,sneider1983children}, (2) gravity \cite{agan2004learning, sneider1998unraveling, nussbaum1976assessment, nussbaum1979children, sneider1983children, ampartzaki2016astronomy}, (3) the day-and-night cycle \cite{chiras2008day, vosniadou1994mental, fleer1997cross}, (4) the seasons \cite{baxter1989children, atwood1996preservice, furuness1989children, trumper2006teaching, sneider2011learning}, and (5) the Earth-Sun-Moon system \cite{wilhelm2018middle, jones1987children, kuccukozer2009effect}. However, there is a scarcity of studies exploring student learning in other areas of astronomy, such as the Big Bang \cite{christonasis2023religiosity,hennig2023big,aretz2016fairytale,trouille2013investigating}, black holes \cite{ubben2022holes}, stars \cite{bailey2003review, bailey2009college}, and aspects related to sizes and distances of astronomical objects \cite{rajpaul2018cross}. A recently published review article by Salimpour \cite{salimpour2022cosmic} further highlights the development of  ``a fertile landscape for research into Cosmology Education'' (p. 819). It is noteworthy that research in the field of astronomy education sketched above has focused on questions regarding the teaching and learning of astronomy concepts in various target groups: These groups range from students, including school-aged students \cite{gali2021secondary, baxter1989children, sadler1992initial, fleer1997cross, rajpaul2018cross}, as well as college and university students \cite{bailey2009college, trumper2000university, berendsen2005conceptual, wallace2011study, trumper2003need}, to pre- and in-service teachers \cite{kalkan2007science,plummer2015preservice, bektacsli2014service, arslan2016pre}.

\subsubsection{Tools to assess students' conceptions of astronomy topics}

To assess students' conceptual understanding and students' conceptions of the aforementioned astronomy topics, a number of concept inventories have been developed and evaluated, e.g., the Lunar Phases Concept Inventory \cite{lindell2002developing}, the Moon Phases Concept Inventory \cite{chastenay2020development}, the Light and Spectroscopy Concept Inventory \cite{bardar2006need,bardar2007development}, the Astronomy Diagnostic Test \cite{hufnagel2000pre, zeilik2002birth,hufnagel2002development}, the Test of Astronomy Standards \cite{slater2014development} or the Star Properties Concept Inventory \cite{bailey2012development}. However, in Ref.~\cite{Bitzenbauer-2022}, the authors emphasize the necessity of developing further instruments that allow for the valid assessment of students' conceptions of more advanced topics such as black holes \cite{ubben2022holes} or the Big Bang \cite{hennig2023big} on the one hand, and stars, especially with regard to aspects beyond their basic properties, on the other. The development of such diagnostic tools seems necessary to address the gap in the existing literature concerning students' understanding and students' conceptions of a broad range of facets related to the topic of stars. In the upcoming subsection~\ref{sec:conceptions_stars}, we will provide an overview of astronomy education research results on this topic.  

\subsection{Students' conceptions of stars}
\label{sec:conceptions_stars}
Agan's study \cite{agan2004stellar} shed light on students' ideas about stars across various educational levels, revealing a multitude of little elaborate ideas: These encompass the twinkling nature of stars or the students' idea that Polaris (the North Star) would be the brightest star. Notably, a significant number of students fail to recognize the Sun as a star, harbor the notion that stars are immortal, and mistakenly assume that all stars end in supernovae. Moreover, an erroneous view persists among students that all stars in the celestial sphere are equidistant from Earth \cite{sharp1996children, sharp1997primary}. Further research has unveiled that students often describe stars as round objects without edges \cite{sharp1996children} and perceive them as motionless entities in the night sky while recognizing the Sun's apparent motion during the day \cite{vosniadou1994mental, slater2015analysis}. Their comprehension of daily celestial motion and knowledge of significant celestial objects, such as Polaris and the ecliptic, also display limitations \cite{dove2002does, plummer2011children}. Additionally, misconceptions regarding the size of stars persist among learners \cite{anantasook2018thai}.

To (a) address these persistent ideas opposing the scientific view and (b) enhance students' learning experiences of astronomy concepts, researchers have explored the impact of out-of-school learning, particularly through planetarium visits. Lelliott \cite{lelliott2007learning} and Dunlop \cite{dunlop2000children} conducted investigations into the effects of planetarium visits on secondary school students' understanding of astronomy topics. Lelliott's study \cite{lelliott2007learning} hints at different cognitive levels of students' knowledge, with planetarium visits leading to changes in their initial ideas and a shift towards more scientifically accurate notions of celestial motion.

\subsection{Research desiderata}
To succinctly summarize, research in astronomy education has revealed that many learners possess limited and fragmented ideas about stars. Prevailing students' ideas opposing the scientific views encompass various aspects such as star formation, size, their relationship to planets, and their position in space. These challenges persist across diverse age groups and educational settings, underscoring the need for further investigation. Specifically, two key research desiderata emerge:
\begin{enumerate}
    \item \textit{Comprehensive analysis of students' views of stars:} The studies presented in subsection~\ref{sec:conceptions_stars} have predominantly focused on isolated aspects of stars, leaving some areas unexplored. Consequently, there is a dearth of instruments capable of capturing students' views comprehensively, encompassing the multifaceted nature of the topic \cite{Bitzenbauer-2022}. A comprehensive analysis and identification of students' views of stars, spanning the various sub-aspects (e.g., from star formation to spectral aspects), remains elusive. Recognizing the prevalent ideas among learners that contradict current scientific views, however, is vital, as it enables the development of targeted instructional interventions. Thus, a pressing research need exists to develop an instrument that can holistically assess learners' views on different aspects related to stars.
    \item \textit{Cross-age exploration of students' learning about stars:} Although existing cross-age studies in astronomy have examined students' learning during limited periods of their school careers \cite{plummer2009cross, hannust2007children}, a comprehensive overview of students' learning about stars across their entire secondary school careers is lacking. Understanding the evolution of students' views of stars over time is crucial for designing instructional strategies that align with their cognitive development and evolving needs. By gaining insights into the progression of students' views throughout their secondary education, educators and curriculum developers can tailor interventions that effectively address conceptual challenges at different stages. 
\end{enumerate}

In this article, we tackle both of these research desiderata: On the one hand, we aim at developing an instrument to economically asses learners' views of different aspects regarding stars. On the other hand, we use this new instrument to explore the development of secondary school students' views of stars throughout their secondary school careers, and hence, aim at gaining a comprehensive overview of widespread students' ideas of stars opposing the current scientific views.

\section{Research Questions}
\label{RQs}
The first research objective of this paper is to develop an instrument that can economically assess learners' views on various aspects of the stars and that performs well psychometrically on a sample of secondary school students. The second research objective is to gain insights into secondary school students' views of stars by examining their continuous development throughout secondary education and identifying widespread ideas that contradict the current scientific views. Hence, we aim at clarifying the following research questions:
\begin{enumerate}[leftmargin=1cm]
\item[RQ1:] How do the proportions of students' ideas about stars aligning with the scientific views compare among lower, middle and upper secondary school students?
\item[RQ2:] What ideas opposing the scientific view on various aspects of stars are widespread among secondary school students?
\end{enumerate}

\section{Design and Evaluation of the Instrument}
\label{sec:design}
In this section, we provide an overview of the development and evaluation of an instrument suitable for the assessment of learners' views of various aspects of stars. This endeavor aligns with the research objective outlined in the previous section (see Section \ref{RQs}) and, hence, contributes to the validity and reliability of the research findings put forth in this paper.

\subsection{Design of the instrument}
\subsubsection{Determination of target group and question format}
Research question 1 addresses the evolution of learners' views of stars throughout their secondary school careers. Hence, the primary target group are secondary school students. To allow for the economic identification of widespread student ideas of stars that are opposing the scientific views on a large scale (see Research question 2), the use of rating scale items seems a sensible approach which has been used in prior research that aimed at the economic assessment of students' views in science education research \cite{muller2002teaching,reinisch2023assessing}. Consequently, in our instrument we decided to include statements about different aspects of stars alongside a four-point rating scale (1 corresponds to "I do not agree", 2 to "I rather do not agree", 3 to "I rather agree" and 4 to "I agree"). We decided to include a response option for abstaining to ensure that participants were not compelled to choose either agreement or disagreement when uncertain. Additionally, this reduces the likelihood of participants guessing their responses (for similar arguments see \cite{veith2022assessing}).

It is crucial to emphasize that the ratings provided by the participants for the statements in the instrument were not assigned as either 'true' or 'false.' Instead, we employed a categorization approach to classify the students' ratings into the following distinct categories: (a) ``in line with the scientific view'', (b) ``opposing the scientific view'', and (c) ``abstained from voting''. This categorization decision is well-justified given the dynamic nature of research on stars and their formation, which continues to grapple with unresolved questions in the field, as thoroughly discussed in the introduction of this article (cf. Section~\ref{sec:intro}). However, we refrain from a more fine-grained categorization that takes into account the degree of (dis)agreement among students, since the meaning assigned to the rating options is highly subjective and may therefore vary from person to person. For example, if a questionnaire item contained a statement that deviated from the currently accepted scientific view of stars, and a student selected the response option 'I rather agree,' we categorized the student's rating as 'opposing the scientific view. 

Lastly, we highlight that we deliberately ask for students' views or students' ideas instead of students' conceptions in the research questions underlying this study. The term conception would refer to representations and notions that people give to phenomena or their underlying patterns \cite{rickheit1999mental}. However, we assume that in this study we also collect ad hoc triggered mental connections of the learners, which are provoked by the items of the research instrument developed in this study, and which would not have been expressed without the input. These are then not internalized conceptions, but we refer to those as students' views or ideas that potentially are more superficial in nature. The distinction of the terms students' views or ideas from the term students' conceptions is in line with diSessa's knowledge-in-pieces perspective on learning (e.g., see \cite{disessa1993ontologies,disessa1998just}) in which students' views or ideas are regarded the ``product of occasional mismatches between p-prims and 
contexts'' (\cite{amin2014student}, p. 10).

\subsubsection{Description of the content domain}
\label{sec:Content}
With our instrument, it should be possible to capture students' understandings of all aspects of stars relevant for astronomy education at the secondary level. Therefore, the aspects to be included were initially identified based on (German) secondary school curricula. Additionally, we used typical undergraduate textbooks (e.g., \cite{spatschek2003astrophysik,hanslmeier2002einfuhrung,scholz2018physik}) as well as scientific articles on astronomy education research that cover student learning (e.g., see \cite{trouille2013investigating,agan2004stellar,adams2000astronomy,vosniadou1992designing,dunlop2000children}) for the description of the relevant content domains. This procedure led to the identification of five thematic domains covering the relevant aspects of stars: 
\begin{itemize}
    \item Domain 1: Stars and solar system. This domain 
    covers the Earth-Sun-Moon relationship as well as their classification as celestial objects. 
    \item Domain 2: Formation and Evolution of Stars. This domain covers topics such as stars' origin, age, life and death.
    \item Domain 3: Properties of Stars. This domain includes key properties of stars, such as size or apparent motion.
    \item Domain 4: (Sub-)Stellar objects. This domain comprises main aspects of binary stars, brown dwarfs, white dwarfs and pulsars.
    \item Domain 5: Spectral aspects. This domain covers questions on stars' specific colors and their brightness. 
\end{itemize}

In the initial iteration, we meticulously devised a comprehensive set of 82 rating scale items, encompassing the five thematic domains (i.e., sub-scales) mentioned above, to be subjected to thorough evaluation (see next Subsection~\ref{sec:Pilot}). These 82 items were partly developed newly, and additionally, we also drew upon earlier instruments and empirical findings gained from prior research.

\subsection{Evaluation of the instrument}
\label{sec:Pilot}

\subsubsection{Expert survey}
To ensure the content validity of the instrument we conducted an expert survey involving a panel of three esteemed faculty members with extensive expertise in astronomy research and teaching. The expert survey encompassed two key aspects: content evaluation and linguistic refinement.

Regarding content evaluation, the experts provided valuable insights on the scientific accuracy of the items and their alignment with the five content domains covered in Section~\ref{sec:Content}. Their comments helped identify any necessary adjustments or reallocations. Simultaneously, the experts scrutinized the language employed in the items, offering recommendations for rephrasing where deemed essential.

Based on the invaluable feedback obtained from the expert survey, we refined the item set, resulting in a total of 65 revised items. These revised items were then subjected to further piloting.

\subsubsection{Psychometric characterization}
\label{subsec:psychometric}
Finally, we conducted a psychometric evaluation of the remaining items. Therefore, the items were administered to a sample of $N = 390$ secondary school and college students. The psychometric characterization of the items based on classical test theory was carried out to select the final item set for the five sub-scales. This involved examining the items' difficulties, with an accepted range of $0.2$ to $0.8$ according to \cite{kline2005psychological}, as well as their discriminatory powers, with accepted values $\geq 0.2$ according to \cite{jorion2015analytic}. Additionally, Cronbach's Alpha was calculated as an estimator of the internal consistency of all five sub-scales~\cite{taber2018use}.

During the assessment, a total of 10 items were identified and subsequently excluded due to their inadequate psychometric characteristics. Consequently, the final instrument comprises 55 items, distributed among the five sub-scales. A comprehensive overview of the item distribution per sub-scale, along with the internal consistencies of each sub-scale, the average item difficulties and discriminatory powers, is given in Table~\ref{Tab:Psycho}. Furthermore, to enhance clarity, we have included a sample item for each sub-scale. The final version of the instrument, with items arranged by sub-scales and alongside potential references, can be found in the appendix of this paper.

Taken together, the rigorous process of development and evaluation culminated in a final version of the instrument that exhibits robust psychometric properties and enables a reliable assessment of secondary school students' views of various aspects of stars. Subsequently, this refined instrument was employed in our main study, which focuses on addressing the research questions outlined in Section~\ref{RQs}. In the following section, we provide a detailed account of the methodology employed in our cross-age study, followed by the presentation of our findings in Section~\ref{sec:results}.

\begin{table*}[hbt]
\vspace*{-\baselineskip}
\caption{\label{Tab:Psycho} Overview of the five sub-scales D1 to D5 (including the number $N$ of items comprised and the Cronbach's $\alpha$ values) of the final version of the instrument used in this study to approach a clarification of the research questions alongside the average item difficulties and the average discriminatory powers of the corresponding items. We refrain from reporting the item difficulties and discriminatory powers of the single items due to the large number of items.}
\begin{ruledtabular}
\begin{tabular}{p{5cm}ccccp{4cm}}
Sub-scale & N & $\alpha$ & $\varnothing$ Item difficulty & $\varnothing$ Discriminatory power & Anchor example\\
\hline	
D1: Stars and solar system  & 9 & $0.69$ & $0.65$ & $0.36$ & The Sun is the largest star in the universe. \\
\hline
D2: Formation and Evolution of Stars  & 16 & $0.83$ & $0.56$ & $0.44$ & Stars do not form and die, they only undergo changes over time. \\
\hline
D3: Properties of Stars & 15 & $0.78$ & $0.66$ & $0.39$ & All stars have the same mass. \\
\hline
D4: (Sub-)Stellar objects & 8 & $0.79$ & $0.35$ & $0.50$ & All stars end as white dwarfs. \\
\hline
D5: Spectral aspects & 7  & $0.73$ & $0.70$ & $0.46$ & All stars are white. 
\end{tabular}
\end{ruledtabular}
\end{table*}

\def\arraystretch{1.3}
\section{Methods}\label{sec:methods}

\subsection{Study design, sample and instrument}
A cross-age study design was chosen to approach a clarification of our research questions as has been done in previous studies concerned with similar research objectives (cf.~\cite{calik-14,trumper-93}). The sample comprised $N=366$ German secondary school students, divided into three different cohorts, enabling a deeper investigation of the temporal progression of students' views of stars throughout secondary education (for limitations of this approach see Section~\ref{sec:limitations}): We included participants from various grades, namely $N_1=148$ ($81$ female) students from grades 7-8 (aged 13-14 years), $N_2=151$ ($70$ female) students from grades 9-10 (aged 15-16 years) and $N_3=67$ ($31$ female) students from grades 11-12 (aged 17-18 years). In the following, we will refer to these subsamples as Cohort 1 (lower secondary school), Cohort~2 (middle secondary school) and Cohort 3 (upper secondary school), respectively. The participants did not receive any instruction as part of this study prior to test administration beyond their regular physics lessons and participation was completely voluntary as well as uncompensated. It it is noteworthy that current physics curricula in Germany deal with astronomy topics in a rather superficial manner, in particular many of the topics assessed by our instrument remain a fringe topic throughout the entirety of secondary education. An overview of the sample is provided in Table~\ref{tab:sample}. 

\begin{table}[H]
\vspace*{-\baselineskip}
\caption{\label{tab:sample} Overview of the sample for the main study.}
\begin{ruledtabular}
\begin{tabular}{cccc}
Cohort & Description & Grades & Sample Size\\
\hline	
1 & lower secondary school & 7-8 & 148\\
2 & middle secondary school & 9-10 & 151\\
3 & upper secondary school & 11-12 & 67
\end{tabular}
\end{ruledtabular}
\end{table}

The data collection, took place at the end of the school year 2021/2022 when the instrument presented in Section~\ref{sec:design} was administered as a paper-pencil-test (for all items of the five sub-scales, see the appendix). The Cronbach Alpha values for the five-sub-scales calculated from the main study data (cf. Table~\ref{tab:Cronbach}) were stable compared to the ones obtained in the pilot study (cf. Section~\ref{subsec:psychometric}).

\begin{table}[H]
\vspace*{-\baselineskip}
\caption{\label{tab:Cronbach} Internal consistencies of the sub-scales of the instrument based on the data of the main study sample. These compare well with the ones obtained in the pilot study (cf. Section~\ref{subsec:psychometric}). For the descriptions of the sub-scales see Section~\ref{sec:Content} and Table~\ref{Tab:Psycho}, respectively.}
\begin{ruledtabular}
\begin{tabular}{ccccc}
          &                & Cronbach's Alpha & \\
Sub-scale & Cohort 1 & Cohort 2         & Cohort 3 \\
\hline	
1 &  0.67 & 0.70 & 0.74\\
2 &  0.86 & 0.87 & 0.79\\
3 & 0.74 & 0.79 & 0.79\\
4 &  0.86 & 0.80 & 0.84\\
5 & 0.71 & 0.73 & 0.71\\
\end{tabular}
\end{ruledtabular}
\end{table}

\subsection{Data analysis}

\subsubsection{Analysis carried out to answer research question 1}
Analyses of variance (ANOVAs) were conducted to check for differences between the three cohorts, with corresponding Tukey-Kramer post-hoc tests to check for significant differences between the groups. As a measure of effect size for the overall comparisons we used partial eta squared ($\eta_p^2$) where the commonly used categorization of small ($\eta_p^2<0.06$), medium ($0.06\leq\eta_p^2<0.14$) and large ($0.14\leq\eta_p^2$) effects was applied~\citep{cohen-1988}. As a measure of effect size regarding the pairwise comparisons we used Cohen's $d$ alongside the established ranges of small ($d<0.5$), medium ($0.5\leq d<0.8$) and large ($0.8\leq d$) effect sizes~\cite{cohen-1988}. To ensure the assumption of homogeneity underlying ANOVA, we employed Levene's Test \cite{levene-1960}. Additionally, the normal distribution of the data was assessed using the Shapiro-Wilk test \cite{shapiro1965analysis}.

\subsubsection{Analysis carried out to answer research question 2}
To analyse the students' views in terms of scientific accuracy we employed the categorization of responses described in Section~\ref{sec:design}, namely (a) ``in line with the scientific view'', (b) ``opposing the scientific view'' and  (c) ``abstained form voting''. In a next step, the proportion of agreements with statements opposing the current scientific views was analysed and an interpretation of the corresponding students' views was given. 

\section{Results}\label{sec:results}
In the following, we report the results of our study, separated by domain. For each of the five investigated domains we will first report ANOVA results to evaluate the development of secondary school students' views of stars. In a second step, we will analyze all responses on the corresponding sub-scale of our instrument to identify widespread ideas opposing scientific views among secondary school learners. We refer to certain items of the instrument with the abbreviation x-y, which stands for item y of sub-scale x (for an overview of all items, see the appendix). 
To provide a concise overview, the ANOVA results are gathered in Table~\ref{tab:anova} and will be explored more thoroughly in the following subsections.


\begin{table*}
\centering
\caption{\label{tab:anova}Results of ANOVAs comparing the percentage of responses aligned with the scientific view on all items across the three cohorts (1 corresponding to lower secondary students, 2 to middle secondary students, and 3 to upper secondary students) and across all sub-scales (domains D1 to D5). The $p$ values reported in the last three columns belong to a Tukey-Kramer post-hoc test. Cohen's $d$ coefficients as measures of effect size for the pairwise comparisons are provided in the corresponding Figures~\ref{fig:box-d1}-\ref{fig:box-d5}.}
\begin{ruledtabular}
\begin{tabular}{cccc>{\centering}p{3cm}ccc|c|c|c}
 &  &  &  & \multirow{2}{3cm}{Sum of squares of residual error} & $F$ & $p$ & $\eta_{p}^{2}$ & \multicolumn{3}{c}{Post-hoc test}\tabularnewline
\cline{9-11} \cline{10-11} \cline{11-11} 
Domain &  & Sum of squares & df &  &  &  &  & 1-2 & 1-3 & 2-3\tabularnewline
\hline 
\hline 
\multirow{2}{*}{D1} & between groups & 3.30 & 2 & 1.65 & 39.7 & $<0.001$ & 0.18 & $<0.01$ & $<0.01$ & $<0.05$\tabularnewline
 & within group & 15.10 & 363 & 0.04 &  &  &  &  &  & \tabularnewline
\hline 
\multirow{2}{*}{D2} & between groups & 2.48 & 2 & 1.24 & 28.0 & $<0.001$ & 0.13 & $<0.01$ & $<0.01$ & $<0.01$\tabularnewline
 & within group & 16.06 & 363 & 0.04 &  &  &  &  &  & \tabularnewline
\hline 
\multirow{2}{*}{D3} & between groups & 1.82 & 2 & 0.91 & 24.0 & $<0.01$ & 0.12 & $<0.01$ & $<0.01$ & $<0.01$\tabularnewline
 & within group & 13.78 & 363 & 0.04 &  &  &  &  &  & \tabularnewline
\hline 
\multirow{2}{*}{D4} & between groups & 0.65 & 2 & 0.32 & 4.46 & $< 0.05$ & 0.02 & $0.94$ & $<0.05$ & $<0.05$\tabularnewline
 & within group & 26.26 & 363 & 0.07 &  &  &  &  &  & \tabularnewline
\hline 
\multirow{2}{*}{D5} & between groups & 1.89 & 2 & 0.94 & 13.2 & $<0.001$ & 0.07 & $<0.05$ & $<0.001$ & $<0.01$\tabularnewline
 & within groups & 25.99 & 363 & 0.07 &  &  &  &  &  &
\end{tabular}
\end{ruledtabular}
\end{table*}

\subsection{Domain 1: Stars and solar system}
Table~\ref{tab:d1-descr} shows the descriptive statistics for all cohorts regarding items of domain 1. While cohort 1 students averaged 48.4\% scientifically accurate answers, the proportion of responses aligning with scientific views for cohort 2 is 64.2\% and further increases to 72.6\% for cohort 3. A similar observation can be made for the median. 

\begin{table}[H]
\vspace*{-\baselineskip}
\caption{\label{tab:d1-descr} Descriptive statistics for the percentage of responses on all items of domain 1 that are in line with the scientific views, separated by cohort.}
\begin{ruledtabular}
\begin{tabular}{cccccc}
Cohort & Mean & SD & Median & Min & Max\\
\hline	
1 & 48.4 & 22.9 & 55.6 & 0.0 & 88.9 \\
\hline
2 & 64.2 & 18.6 & 66.7 & 11.1 & 100 \\
\hline
3 & 72.6 & 18.3 & 77.8 & 22.2 & 100 
\end{tabular}
\end{ruledtabular}
\end{table}

\subsubsection{ANOVA results for domain 1}
\begin{figure}
     \centering
     \includegraphics[width=\linewidth]{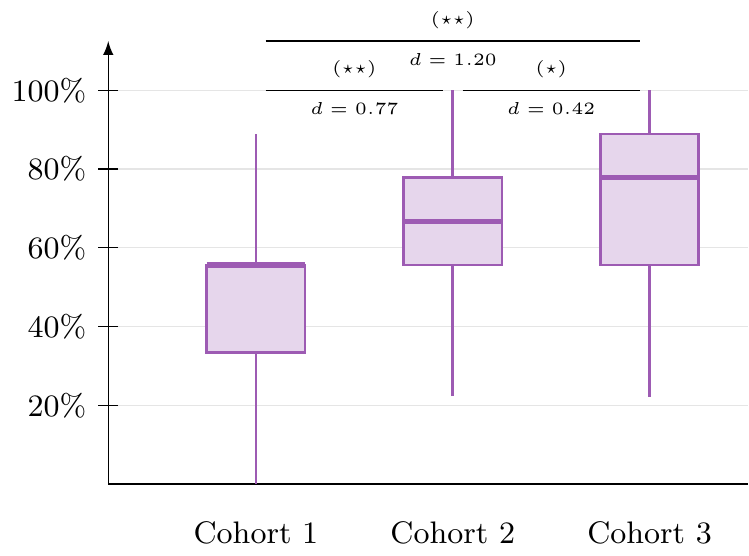}
     \caption{Boxplot for the percentage of responses on all items of domain 1 that are in line with the scientific views. Asterisks indicate the statistical
significance of Tukey-Kramer post hoc pairwise comparisons (*: $p < 0.05$, **: $p < 0.01$, ***:~$p~<~0.001$), whereas Cohen's $d$ is reported as a measure of effect size. }
     \label{fig:box-d1}
\end{figure}

The trend observed in Table~\ref{tab:d1-descr} is statistically substantiated by the ANOVA results (cf. Table~\ref{tab:anova}). The difference between the three cohorts is statistically significant [$F(2,363)=39.7, p<0.001; \eta_p^2=0.18$]. Comparing the three cohorts directly yields a statistically significant difference between cohort 2 and cohort 3 $(p<0.05)$ with a effect size of $d=0.42$. Cohort 1, on the other hand, differs highly significantly from both cohort 2 ($p<0.01$) and cohort 3 ($p<0.01$) with medium to high effect sizes of $d=0.52$ and $d=1.07$, respectively. These results are summarized in the form of boxplots in Figure~\ref{fig:box-d1}  -- for the presentation of our boxplots the whiskers indicate $1.5\times \operatorname{IQR}$ throughout this article, where IQR is the interquartile range.

\subsubsection{Secondary school students' views of stars and the solar system}
A more in-depth view of the student's responses on the items of domain 1 is provided by Table~\ref{tab:d1}. Here, we report the more fine-grained categorization of answers into answers that are in line with scientific views ($+$), opposing scientific views ($-$) and the ones abstained from voting ($\circ$). Across all items and cohorts, there is generally a high percentage of students providing scientifically sound answers. Nevertheless, the ANOVA results are also reflected in the observation that the share of answers in line with the scientific view increase from throughout grades 7 to 12. An item that stands out due to a large share of $-$ is item 1-4, indicating that between 32.4\% (lower), 46.1\% (middle) and 41.8\% (upper) secondary school students believe that ``the planets and the Sun were formed at the time of the Big Bang''. An even higher share of answers opposing the scientific view and the most evident view held by all students is revealed by item 1-5 which states that ``there are hundreds of stars in our solar system.'' Here, 75.0\% (lower), 70.9\% (middle) and 62.7\% (upper) of all participants agreed with the statement, indicating trouble with grasping the magnitude of our solar system. Lastly, item 1-6 exhibits the greatest share of abstained votes, with 32.4\% (lower), 25.8\% (middle) and 28.4\% (upper), respectively. In other words, this item stating that ``metals have existed in the universe since the Big Bang'' was met with the highest uncertainty and, consequently, a low percentage of scientifically sound responses.

\subsection{Domain 2: Formation and evolution of stars}
Table~\ref{tab:d2-descr} shows the descriptive statistics for all cohorts regarding items of domain 2. While mean and median are, on average, slightly lower compared to domain 1, we again observe an increase of all metrics with the exception of standard deviation.

\begin{table}[H]
\vspace*{-\baselineskip}
\caption{\label{tab:d2-descr} Descriptive statistics for the percentage of responses on all items of domain 2 that are in line with the scientific views, separated by cohort.}
\begin{ruledtabular}
\begin{tabular}{cccccc}
Cohort & Mean & SD & Median & Min & Max\\
\hline	
1 & 42.9 & 21.4 & 43.8 & 0.0 & 75.0 \\
\hline
2 & 53.8 & 20.8 & 50.0 & 6.3 & 100 \\
\hline
3 & 65.4 & 20.9 & 68.8 & 6.3 & 100 
\end{tabular}
\end{ruledtabular}
\end{table}

\subsubsection{ANOVA results for domain 2}
\begin{figure}
     \centering
     \includegraphics[width=\linewidth]{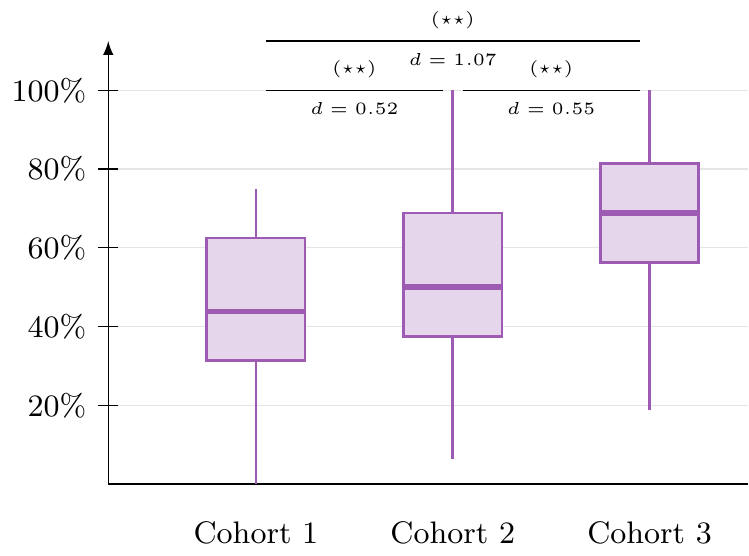}
     \caption{Boxplot for the percentage of responses on all items of domain 2 that are in line with the scientific views. Asterisks indicate the statistical
significance of Tukey-Kramer post hoc pairwise comparisons (*: $p < 0.05$, **: $p < 0.01$, ***:~$p~<~0.001$), whereas Cohen's $d$ is reported as a measure of effect size.}
     \label{fig:box-d2}
\end{figure}

The trend observed for domain 1 also holds for domain 2 (cf. Table~\ref{tab:anova}): The difference between the cohorts is very highly significant [$F(2,363)=28.0, p<0.001; \eta_p^2=0.13$]. Likewise, all between group comparisons show highly statistical significance ($p<0.01$) with effect sizes ranging from $d=0.52$ between cohort 1 and 2, $d=0.55$ between cohort 2 and 3 as well as $d=1.07$ between cohort 1 and 3. Hence, not only do the groups differ highly significantly, but medium to high effect sizes can be associated with each of the pairwise comparisons, indicating a steady increase of scientifically accurate views across secondary education. These results are summarized in the form of boxplots in Figure~\ref{fig:box-d2}.

\subsubsection{Secondary school students' views of the formation and evolution of stars}
A more fine-grained overview of responses is provided in Table~\ref{tab:d2}. Again, we observe a general tendency towards answers in line with the scientific view compared to answers opposing it. Additionally, the pattern of upper secondary school students outperforming their middle and lower peers repeats. An item with a notable percentage of scientifically inaccurate views is item 2-10 which deals with the color of stars. Here, 50.7\% of lower, 49.0\% of middle and 38.8\% of upper secondary school students hold the opinion that stars do not change their color throughout their life, neglecting their evolutionary stages that impact the perceived color. This lack of awareness regarding the evolutionary stages is substantiated by item 2-13 which states that ``stars fade and disappear over time''. An almost equal share of 63.5\% (lower), 58.3\% (middle) and 53.7\% (upper) disagreed with this statement and, thus, perceive stars as permanent objects. Lastly, item 2-16 reveals another view that falls in line with the responses on item 1-5: Almost half of each cohort responded that ``a supernova immediately destroys a large part of the galaxy'', indicating yet again a skewed perception of astronomic scales.

\subsection{Domain 3: Properties of Stars}
The descriptive statistics for all cohorts regarding items of domain 3 are provided in Table~\ref{tab:d3-descr}. The values overall compare very well to the ones from domain 1 and 2 (cf. Table~\ref{tab:d1-descr} and Table~\ref{tab:d2-descr}). 

\begin{table}[H]
\vspace*{-\baselineskip}
\caption{\label{tab:d3-descr} Descriptive statistics for the percentage of responses on all items of domain 3 that are in line with the scientific views, separated by cohort.}
\begin{ruledtabular}
\begin{tabular}{cccccc}
Cohort & Mean & SD & Median & Min & Max\\
\hline	
1 & 55.7 & 22.2 & 60.0 & 0.0 & 93.3 \\
\hline
2 & 64.2 & 17.5 & 66.7 & 13.3 & 100 \\
\hline
3 & 75.3 & 17.2 & 80.0 & 20.0 & 100 
\end{tabular}
\end{ruledtabular}
\end{table}

\subsubsection{ANOVA results for domain 3}
\begin{figure}
     \centering
     \includegraphics[width=\linewidth]{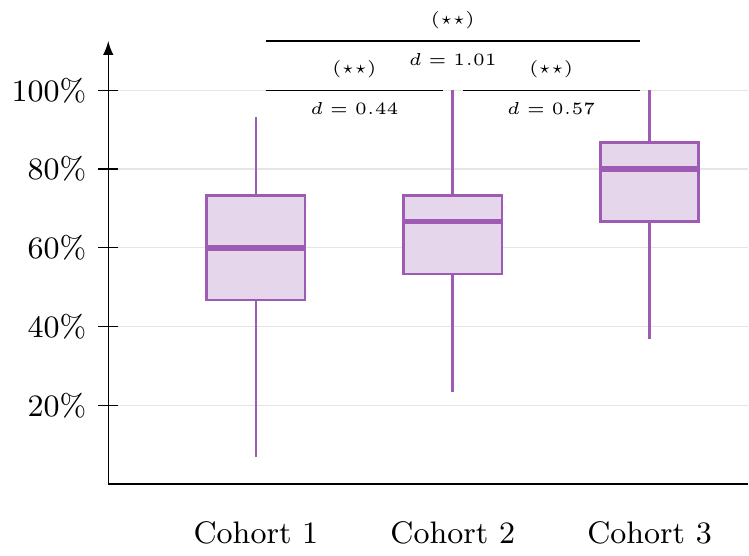}
     \caption{Boxplot for the percentage of responses on all items of domain 3 that are in line with the scientific views. Asterisks indicate the statistical
significance of Tukey-Kramer post hoc pairwise comparisons (*: $p < 0.05$, **: $p < 0.01$, ***:~$p~<~0.001$), whereas Cohen's $d$ is reported as a measure of effect size.}
     \label{fig:box-d3}
\end{figure}

The ANOVA results for domain 3 show almost identical values with those of domain 2 (cf. Table~\ref{tab:anova}). The cohorts differ highly statistically significantly [$F(2,363)=24.0, p<0.01; \eta_p^2=0.12$] and the same is true for all pairwise cohort comparisons ($p<0.01$ each). The corresponding effect sizes are all high with $d=0.44$ between Cohort 1 and 2, $d=0.57$ between Cohort 2 and 3 as well as $d=1.01$ between Cohort 1 and 3. Hence, as before, a continuous improvement in scientifically accurate views held by students can be observed from lower to middle and, lastly, higher secondary education. A summary of these results in terms of Boxplots is provided in Figure~\ref{fig:box-d3}.

\subsubsection{Secondary school students' views of properties of stars}
Table~\ref{tab:d3} provides an overview of the responses on all items of domain 3, exhibiting similar answer patterns as before (cf. Table~\ref{tab:d1} and Table~\ref{tab:d2}). Here, item 3-4 stands out as it extends an idea observed in domain 2: 43.9\% of lower, 29.1\% of middle and 37.3\% of upper secondary school students held the misguided view that stars do not underlie gravitational pull, viewing stars as stationary celestial objects that do not interact with their surroundings. The share of scientifically accurate responses on this item being the same for cohorts 2 and 3 suggests that this view is somewhat persistent throughout higher secondary education. This is further substantiated by the response pattern of item 3-13 which states that ``stars seem to rise and set''. Again, with 35.8\%, 43\% and 34.3\%, respectively, a substantial share of participants disagreed with that statement, expressing yet another view of \textit{rigid} stars. On a different note, item 3-7 reveals that 39.9\% of lower, 42.4\% of middle and 28.4\% of higher secondary school students relate a star's distance to its apparent brightness in the sky, stating that ``the brightest stars are closest to Earth''.

\subsection{Domain 4: (Sub-)Stellar objects}
Table~\ref{tab:d4-descr} provides an overview of descriptive statistics for all items of domain 4. This domain holds the overall lowest descriptive statistics. 
While cohort 1 students averaged 32.9\% scientifically accurate answers, the proportion of responses aligning with scientific views for cohort 2 is 31.8\% and 43.1\% for cohort 3. The median reflects this performance and, for the first and only time, there are participants with no responses that are in line with the scientific view. Thus, on average secondary school students' views differed the most from the current scientific view regarding (sub-)stellar objects. We will elaborate on this observation in the discussion section (cf. Section~\ref{sec:discussion}).

\begin{table}[H]
\vspace*{-\baselineskip}
\caption{\label{tab:d4-descr} Descriptive statistics for the percentage of responses on all items of domain 4 that are in line with the scientific views, separated by cohort.}
\begin{ruledtabular}
\begin{tabular}{cccccc}
Cohort & Mean & SD & Median & Min & Max\\
\hline	
1 & 32.9 & 27.9 & 37.5 & 0.0 & 87.5 \\
\hline
2 & 31.8 & 26.4 & 25.0 & 0.0 & 100 \\
\hline
3 & 43.1 & 25.7 & 37.5 & 0.0 & 100 
\end{tabular}
\end{ruledtabular}
\end{table}

\subsubsection{ANOVA results for domain 4}
\begin{figure}
     \centering
     \includegraphics[width=\linewidth]{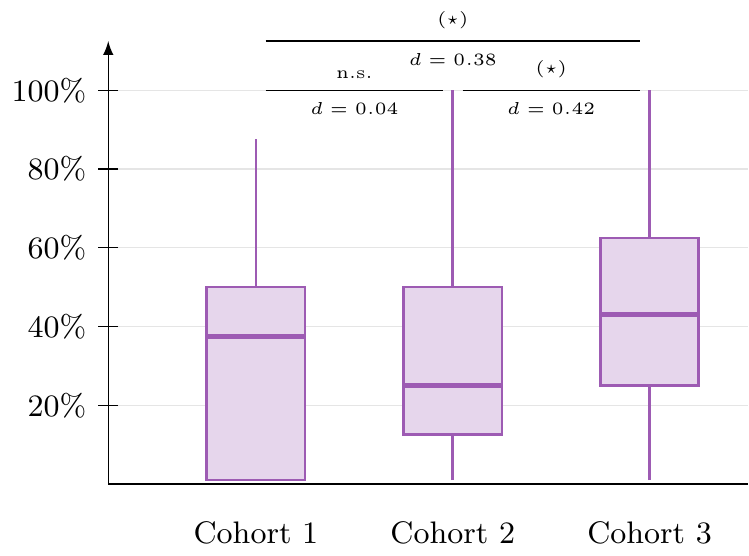}
     \caption{Boxplot for the percentage of responses on all items of domain 4 that are in line with the scientific views. Asterisks indicate the statistical
significance of Tukey-Kramer post hoc pairwise comparisons (*: $p < 0.05$, **: $p < 0.01$, ***:~$p~<~0.001$), whereas Cohen's $d$ is reported as a measure of effect size.}
     \label{fig:box-d4}
\end{figure}

With the scores being relatively low in each cohort, the variance in performance did not differ as much compared to the other domains. While overall group comparison still results in a statistically significant difference [$F(2,363)=4.46, p<0.05;\eta_p^2=0.02$], no statistically significant difference was found between cohorts 1 and 2 ($p=0.94$). The differences between cohort 1 and 3 as well as cohort 2 and 3 were found to be statistically significant by Tukey-Kramer post-hoc tests ($p<0.05$ each) but the corresponding effect sizes were low with $d=0.38$ and $d=0.42$, respectively. Thus, for the domain of (sub-)stellar  objects we record the overall least distinguishable progress throughout secondary education, with only the later grades (11 and 12) meaningfully separating themselves from the rest. The boxplots reflecting this observation are presented in Figure~\ref{fig:box-d4}.

\subsubsection{Secondary school students' views of (sub-)stellar objects}
A deeper look into this evidently more elusive domain is enabled by the response pattern on all items, provided in Table~\ref{tab:d4}. It becomes clear that students on average did not hold inaccurate views but rather abstained from voting, perhaps due to the (currently) marginal nature of the corresponding contents in secondary curricula. Items 4-3, 4-5 as well as 4-7 all address brown and white dwarfs and have in common that students across all cohorts held similar views. With on average one third of all participants expressing views opposing the scientific view across these items, it is suggested that (sub-)stellar objects constitute a fringe topic for which little progress is recorded throughout the grades 7 to 12. 

\subsection{Domain 5: Spectral aspects}
Table~\ref{tab:d5-descr} shows the descriptive statistics for all cohorts regarding items of domain 5. In contrast to domain 4, domain 5 records the overall highest mean percentages of responses in accordance with the scientific views with 58.9\% for lower, 67.3\% for middle and 78.9\% for higher secondary school students. The minimum and maximum in each cohort are 0.0\% and 100.0\%, respectively.

\begin{table}[H]
\vspace*{-\baselineskip}
\caption{\label{tab:d5-descr} Descriptive statistics for the percentage of responses on all items of domain 5 that are in line with the scientific views, separated by cohort.}
\begin{ruledtabular}
\begin{tabular}{cccccc}
Cohort & Mean & SD & Median & Min & Max\\
\hline	
1 & 58.9 & 29.5 & 71.4 & 0.0 & 100 \\
\hline
2 & 67.3 & 24.9 & 71.4 & 0.0 & 100 \\
\hline
3 & 78.9 & 24.3 & 85.7 & 0.0 & 100 
\end{tabular}
\end{ruledtabular}
\end{table}

\subsubsection{ANOVA results for domain 5}
\begin{figure}
     \centering
     \includegraphics[width=\linewidth]{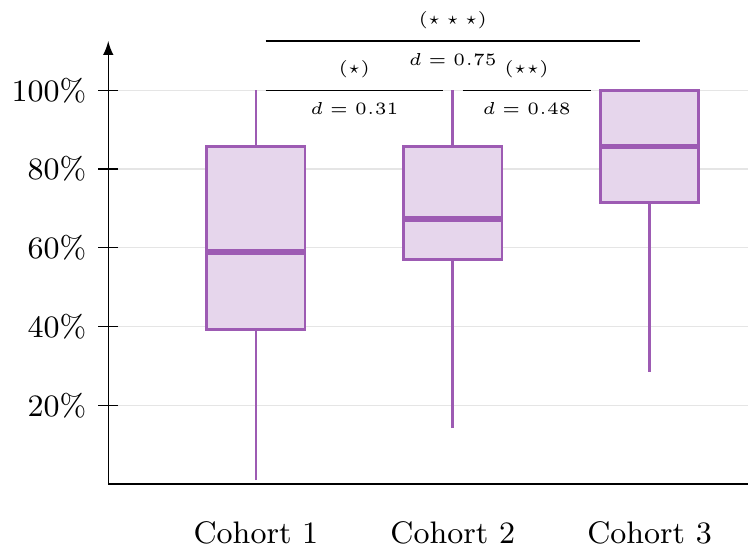}
     \caption{Boxplot for the percentage of responses on all items of domain 5 that are in line with the scientific views. Asterisks indicate the statistical
significance of Tukey-Kramer post hoc pairwise comparisons (*: $p < 0.05$, **: $p < 0.01$, ***:~$p~<~0.001$), whereas Cohen's $d$ is reported as a measure of effect size.}
     \label{fig:box-d5}
\end{figure}

The ANOVA results for domain 5 demonstrate a much pronounced improvement along the trajectory of secondary education (cf. Table~\ref{tab:anova}). The overall difference between the three cohorts is highly statistically significant [$F(2,363)=13.2, p<0.001; \eta_p^2=0.07$]. A Tukey-Kramer post-hoc test indicates that the advancement from grades 7-8 to grades 9-10 is statistically significant ($p<0.05$) with an effect size of $d=0.31$, whereas the difference between grades 9-10 and 11-12 is highly statistically significant ($p<0.01$) with an effect size $d=0.48$. Lastly, the difference between cohorts 1 and 3 is very highly statistically significant ($p<0.001$) with a medium effect size of $d=0.75$. The boxplots reflecting this continuous increase in accurate scientific views are illustrated in Figure~\ref{fig:box-d5}. 

\subsubsection{Secondary school students' views of Spectral aspects}
The fine-grained overview of responses to items of the final sub-scale is presented in Table~\ref{tab:d5}. Here, the item that stands out the most is item 5-7 which states that ``it is said that stars twinkle because they change their brightness'' -- 45.9\% of lower, 39.1\% of middle and 25.4\% of higher secondary school students agreed with this statement. In congruence to the response pattern on items 2-10 and 3-7 singled out earlier, the brightness of stars seems to constitute a recurring theme where students' views are inaccurate to a greater extent than usual. We will discuss this observation more thoroughly in Section~\ref{sec:discussion}.

\section{Discussion}\label{sec:discussion}
In this section, we contextualize our findings against the backdrop of prior astronomy education research. In particular, we shed light on the extent to which our findings are consistent with prior research, identify areas where differences have emerged, and highlight new contributions to the understanding of secondary school students' views of stars. Moreover, we discuss implications of our findings for both astronomy education research and practice.

In general, our cross-age analysis reveals a progressive development of students' perspectives on star-related topics throughout their secondary school education. The statistically significant increase in the percentage of responses aligned with current scientific views from lower to upper secondary school applies to all aspects of stars examined in this study, as demonstrated through ANOVAs and post-hoc pairwise comparisons (see Section \ref{sec:results}). Furthermore, our study expands upon existing literature, which primarily focuses on students' views of the nature, apparent motion, and properties of stars \cite{agan2004learning, plummer2009cross, bailey2012development}. 

\subsection{Discussion of findings regarding domain 1}
The percentage of responses aligning with scientific views in domain 1 (stars and solar systems) shows an increase from 48.4\% among lower, and 64.2\% among middle to 72.6\% among upper level secondary school students (see Table \ref{tab:d1-descr}). An item with a very big gain of alignment towards scientific views was for example item 1-9, where progressively more students disagreed with the statement that the Sun is the largest star in the universe (75.0\% of the lower, 43.7\% of the middle and 34.3\% of the upper secondary students).
On the single-item level we also found, that a majority of the study participants (75.0\% of the lower, 70.9\% of the middle and 62.7\% of the upper secondary students) agreed that there are hundreds of stars in our solar system (item 1-5, see Table~\ref{tab:d1}). This idea opposes the scientific view and has already been reported in Ref.~\cite{comins2001heavenly}. This student view can possibly be explained by the confusion of the terms solar system and stellar system as indicated by Rajpaul et al.~\cite{rajpaul2018cross}. Another widespread perspective among our study participants is that metals have existed since the Big Bang (item 1-6, see Table~\ref{tab:d1}): 46.6\% of the lower, 40.4\% of the middle, and 29.9\% of the upper secondary students were of that opinion, while in addition, around a third of the students at all grade levels were undecided (32.4\%, 25.8\% and 28.4\%, respectively). This finding supports the study by Slater et al. \cite{slater2015analysis}, in which around a third of the students have been reported to hold the view that heavy atoms have existed since the Big Bang. 

In contrast to previous research by Dunlop~\cite{dunlop2000children}, Philips~\cite{philips1991earth} or Comins~\cite{comins2001heavenly}, in our study, we did not find evidence of students misconceiving that the Sun is not a star: Only approximately 10\% of lower and middle secondary school students, and merely 3\% of upper secondary school students, subscribed to this view (see see item 1-8 Table~\ref{tab:d1}). It is noteworthy, that it remains unclear which research methods Philips \cite{philips1991earth} and Comins \cite{comins2001heavenly} employed to uncover this confusion to be widespread among learners. Additionally, Comins \cite{comins2001heavenly} stated that many students would believe that the sun is bigger than other stars. In our study, we identified this view to be widespread among the lower secondary school students (60.1\%, see Table~\ref{tab:d1}), while only 37.7\% of the middle and 29.9\% of the upper secondary students agreed with item 1-9 (``The sun is the largest star in the universe.''). 


\subsection{Discussion of findings regarding domain 2}
Similar to domain 1, the percentage of responses aligning with scientific views in domain 2 (formation and evolution of stars) shows an increase from 42.9\% among lower secondary school students to 53.8\% among middle secondary school students and 65.4\% among upper secondary school students (see Table \ref{tab:d2-descr}). An item with a very big gain of alignment towards scientific views was, for example, item 2-10, where progressively more students agreed that stars can change in color (16.9\% of the lower, 35.8\% of the middle and 79.1\% of the upper secondary students). But on the single-item level, we also found that -- in line with earlier research conducted by Agan \cite{agan2004stellar} -- more than half of our participating students agreed with item 2-13 stating that stars would fade and disappear over time (63.5\% of the lower, 58.3\% of the middle, and 53.7\% of the upper secondary students). 

\subsection{Discussion of the findings regarding domain 3}
\label{sec:diss_3}
A study similar to the one presented in this paper, focusing on star-related aspects, was conducted by Plummer \cite{plummer2009cross}. In her cross-age study, Plummer aimed to assess students' views of celestial motion and identify any misconceptions held at different grade levels (1st, 3rd, and 8th grade students). One part of her study specifically focused on students' views of the apparent motion of stars. Among the sample students, two main perspectives emerged: those who could provide a general description of stars moving slowly across the night sky (40\% of 8th graders, 50\% of 3rd graders, and 35\% of 1st graders), and those who believed that stars never move (40\%, 40\%, and 25\%, respectively). The remaining students held various perspectives, including the idea that stars only move at the end of the night. Our study complements Plummer's findings by providing insights into the views of older students regarding the apparent motion of stars. Among our cohort 1 students (grades 7 and 8), only 18.2\% believed that stars are stationary and fixed in the sky (item 3-11, see Table~\ref{tab:d3}). This percentage was even lower among cohort 3 students (grades 11 and 12) at 14.9\%. In addition, our findings align with the research conducted by Agan \cite{agan2004stellar}, revealing a comparable proportion of students (approximately 14\%) who share the view that stars are stationary. This stands in contrast to studies published in Refs.~\cite{vosniadou1994mental, slater2015analysis, sadler1992initial}, which reported a higher prevalence of this view among approximately 40\% of students. On a further note, our findings are supported by Plummer's study, according to which more than ``one-half of the students in the first, third, and eight grades do not think that we see different stars in the sky during the night (65\%, 60\%, and 65\%), respectively'' (\cite{plummer2009cross}, p. 1598). In our study, 60.8\% of lower secondary school students, 75.5\% of middle secondary school students, and 82.1\% of upper secondary school students agreed with item 3-12, which states that we see different stars over the course of a night (see Table~\ref{tab:d3}). In summary, our findings suggest a continued evolution of students' views on the apparent motion of stars throughout secondary education. 

Another cross-age study on basic astronomy concepts was conducted by Trumper, which included both junior (grades 7-9) \cite{trumper2001junior} and senior (grades 10-12) \cite{trumper2001senior} high school students. In Trumper's study, only 36\% of junior high school students were aware that stars are the farthest objects from the Earth (\cite{trumper2001junior}, p. 1117), while this percentage increased to 49\% among senior high school students (\cite{trumper2001senior}, p. 103). In our study, the percentages of students agreeing with the scientific views on this matter were slightly higher in the sample cohorts. Among lower secondary school students, 43.9\% agreed that stars are farther away from the Earth than the Sun (item 3-9, see Table~\ref{tab:d3}). This percentage increased to 51.7\% among middle secondary school students and 53.7\% among upper secondary school students. Therefore, it appears that students' views regarding distances develop throughout their school careers. This observation is further supported by the decreasing agreement with item 3-10, which states that the distance between stars is about the same as the distance between planets. While half of the lower secondary school students disagreed with this item, the disagreement percentage increased to 62.9\% among middle secondary school students and 74.6\% among upper secondary school students (see Table~\ref{tab:d3}). Our findings align with earlier research on learners' views of astronomical object distances from the Earth's surface (e.g., see \cite{rajpaul2018cross,hennig2023big,sadler1992initial}).


\subsection{Discussion of findings regarding domains 4 \& 5}
While the percentage of responses aligning with scientific views in domain 4 remains below 50\% for students across all grade levels (see Section \ref{tab:d4-descr}), we found a majority of students demonstrating views in line with current scientific understandings of the spectral aspects of stars in domain 5 (see Section \ref{tab:d5-descr}). 
No statistically significant difference can be observed in the percentage of responses aligning with scientific views for items in domain 4 between lower (32.9\%) and middle secondary school students (31.8\%), see Table~\ref{tab:d4-descr}. This finding can likely be attributed to the omission of (sub-)stellar objects from the German astronomy curricula at both lower and middle school levels. Insights into current curriculum developments regarding astronomy in German secondary schools can be found in Ref.~\cite{hennig2023big}. The topic of spectral aspects shows a similar pattern: Surprisingly, a small but statistically significant difference exists in the percentage of responses aligning with scientific views for items in domain 5 between lower (58.9\%) and middle secondary school students (67.3\%). This discrepancy can possibly be attributed to either implicit learning, which occurs unconsciously as students engage with different topics, such as atomic physics (for more details on implicit learning, see Reber \cite{reber1989implicit}), or the influence of informal learning environments \cite{lelliott2010concept}, such as planetariums \cite{dunlop2000children}. There were items like 4-6 (see Table~\ref{tab:d4}) and 5-6 (see Table~\ref{tab:d5}), where a bigger shift towards a perspective aligned with scientific views was observed (38.5\% to 70.1\% and 39.2\% to 86.6\% respectively), though for most, the shifts remained smaller.
In future research, it would be valuable to explore (a) which topics in the secondary school curriculum facilitate implicit learning of astronomy and (b) the sources from which students gain insights into astronomy topics in informal settings.

\subsection{Implications for Educational Research}
The results of this study imply that, in general, students' ideas about stars show a progressive alignment with the current scientific views. This quite positive development, however, lacks a clear explanation at this point, meaning that further extensive research is advised to clear up the various causes for the developments presented in this study: While teaching in the classroom appears to play a role, it is important to consider the potential influence of informal learning environments such as out-of-school visits \citep{andre2017museums,behrendt2014review} or explanatory videos \citep{kulgemeyer2020framework}. These supplementary resources may contribute to the positive evolution of students' understanding. Additionally, no data about the traits of the students have been gathered. As previous research has shown, astronomical topics tend to rank highly in studies on interesting topics for students \citep{holstermann2007interesse}. A thorough investigation into how and what affective traits facilitate the increase in alignment with scientific topics during the secondary school years could give new and more detailed insights into the role of interests in learning processes. 
Furthermore, the instrument utilized in this study has the potential to be employed in future research to identify the specific factors and properties of learning materials and methodologies that facilitate these favorable learning outcomes. By using this instrument, researchers can gain insights into the effective strategies and approaches that support students' development of ideas in the domain of stars and related astronomical concepts, complementing previous instruments devised for this purpose (e.g., see \cite{bailey2006development, lindell2002developing}).
Further investigation is warranted to delve deeper into the mechanisms driving this positive learning trajectory. Additionally, exploring the comparative impact of various educational interventions and materials could shed light on the most effective approaches for promoting accurate scientific understanding of stars among students. 

\subsection{Implications for Educational Practice}
From the findings presented in this study, we surmise that current educational practices available between the lower and upper secondary school level likely facilitate a basic development of ideas about stars that in general align progressively more with scientific views during these years. However, some persistent ideas have also been isolated that have also been found in corresponding literature (e.g., see \cite{plummer2009cross,agan2004stellar,trumper2001cross}). Mainly, static ideas have been found especially in the lower high school classes and confusions of central ideas like stellar system and solar system have also been documented. Ideas of change  and dynamics (stars changing brightness or stars changing color) might be better facilitated by incorporating learning materials that depict these dynamic properties (such as videos or comics, cf. \cite{kulgemeyer2020framework,wallner2020using}) or even enable learners to interact with the material (such as simulations, cf. \cite{rutten2012learning}).

\section{Limitations}\label{sec:limitations}
While our study on students' views of stars throughout their secondary school careers provides valuable insights, it is important to acknowledge several limitations. Firstly, it is crucial to note that our research adopted a cross-age study design rather than a conventional longitudinal approach: Instead of longitudinally tracking the views of a specific group of students, we assessed students from different grade levels at a single time point. Our approach offers valuable cross-sectional data but (a) limits our ability to capture individual students' developmental trajectories and the specific changes in their views of stars over time and (b) may be subject to cohort bias \cite{desrosiers1998comparison}. Secondly, it is important to acknowledge the limited emphasis on astronomy within the secondary school curriculum in Germany where this study was conducted (e.g., see \cite{falcke1999astronomy}). The lack of continued instruction in astronomy raises the question of how students' views of stars might be influenced if they had participated in continued astronomy instruction throughout their secondary school careers. Thirdly, the subsamples of lower and middle secondary school students consisted of approximately 150 participants each, whereas the number of participants in the upper secondary school group was only about half that size. This discrepancy arises from the fact that in Germany, students make a decision after completing grade 10 regarding whether to pursue a physics course in upper secondary school. Consequently, the total cohort of possible study participants  becomes significantly smaller at this grade level. Furthermore, teachers are often less inclined to participate in research studies during the twelfth grade due to the impending final exams. These factors imposed limitations on our sampling approach, resulting in an asymmetric distribution of participants among the three subsamples.

Another limitation stems from the question format used in our instrument, which consisted of closed-ended rating-scale items. While this format facilitated efficient data collection on students' views of stars on a large scale, it is important to recognize that the predetermined statements may have influenced participants' responses, potentially leading to the generation of ad hoc conceptions. To mitigate this limitation, future research should incorporate qualitative data collection methods, such as mind mapping or concept mapping \cite{deeb2015nitarp,bizimana2022fostering,malatjie2019exploring,mainali2022investigating}, to gather in-depth insights that validate and expand upon our findings. Furthermore, it is crucial to acknowledge that our study primarily focused on assessing students' ideas and views of stars, rather than their conceptual understanding of relevant aspects related to stars. While our study provides valuable perspectives, it is essential to complement these findings with investigations into students' conceptual understanding. Developing a concept inventory should therefore be a priority for future research, enabling the assessment of students' conceptual understanding of the topic under investigation (cf. \cite{Bitzenbauer-2022}).

\section{Conclusion}\label{sec:conclusion}
From the data collected and analyzed, our findings about students' ideas about stars compared to scientific views have shown that progressing from lower to upper secondary school, ideas start to align more with the current scientific views. This development in itself is positive, though the exact factors for it still need to be further determined. From the data gathered, some of the ideas are already being developed rather effectively, such as ideas about Sun's size when compared to other stars (see items 1-9 in Table~\ref{tab:d1} and 3-2 in Table~\ref{tab:d3}) or that stars can change both color and brightness (see items 2-10 in Table~\ref{tab:d2} and 5-6 in Table~\ref{tab:d5}) while others seem more robust, such as the idea that there are hundreds of stars in the solar system (item 1-5, see Table~\ref{tab:d1}) or that white dwarfs are suns (item 4-5, see Table~\ref{tab:d4}). In summary, our analysis indicates a positive trend of students' ideas aligning more closely with scientific views regarding stars as they progress through secondary school, although the specific contributing factors remain to be determined.

\section*{Data availability}
Anonymized data from the study is available on request from the authors. 

\bibliography{output.bib}

\newpage
\appendix

\phantomsection
\section*{Appendix: The instrument used in this study\label{appendix}}
In this appendix, we provide all items of the instrument used in this study, sorted by the five sub-scales included.

\begin{table*}[hbt]
\subsection{Sub-scale on domain 1: Stars and solar system} \label{instrument:domain1}
\vspace{-20pt}
\caption{\label{Tab:RatingScale1} Overview of the rating scale items related to the sub-scale on domain 1. The table includes the item numbers (No.), the statements themselves, indications of whether the statements align with the scientific view or oppose it, and any references or sources of inspiration.}
\begin{tabularx}{\textwidth}{p{0.03\textwidth}p{0.5\textwidth}p{0.15\textwidth}p{0.15\textwidth}p{0.1\textwidth}}
\toprule
No. & Item & In line with scientific view & Opposing scientific view & Reference / Inspired by \\
\midrule	
1-1 & The Earth orbits the Sun and the Moon. & & x & \\
1-2 & Sun and Moon orbit the Earth. & & x & \\
1-3 & Earth and Moon orbit the Sun. & x & & \\
1-4 & The planets and the Sun were formed at the time of the Big Bang.  & & x & \cite{trouille2013investigating}\\
1-5 & There are hundreds of stars in our solar system. & & x & \cite{adams2000astronomy,agan2004stellar}\\
1-6 & Metals have existed in the universe since the Big Bang. & & x & \\
1-7 & The Moon is a star.  & & x & \cite{vosniadou1992designing}\\
1-8 & The Sun is a star.  & x & & \cite{dunlop2000children}\\
1-9 & The Sun is the largest star in the universe.  & & x & \cite{agan2004stellar} \\
\bottomrule
\end{tabularx}
\end{table*}

\begin{table*}[hbt]
\subsection{Sub-scale on domain 2: Formation and
Evolution of Stars} \label{instrument:domain2}
\vspace{-20pt}
\caption{\label{Tab:RatingScale2} Overview of the rating scale items related to the sub-scale on domain 2. The table includes the item numbers (No.), the statements themselves, indications of whether the statements align with the scientific view or oppose it, and any references or sources of inspiration.}
\begin{tabularx}{\textwidth}{p{0.05\textwidth}p{0.5\textwidth}p{0.15\textwidth}p{0.15\textwidth}p{0.1\textwidth}}
\toprule
No. & Item & In line with scientific view & Opposing scientific view & Reference / Inspired by \\
\midrule	
2-1 & Stars already existed before the Big Bang.  & & x & \cite{trouille2013investigating,adams2000astronomy} \\
2-2 & Stars don’t form and die, they just change over time. & & x & \\
2-3 & All stars are orbited by planets. &  & x & \\
2-4 & Stars have spikes.   & & x &\\
2-5 & The surface of stars is almost completely covered by volcanoes.  & & x & \cite{agan2004stellar}\\
2-6 & All stars are the same age. & & x & \\
2-7 & We see the stars exactly as they are in this moment. & & x & \\
2-8 & Stars undergo changes.  & x & & \\
2-9 & Stars change their size.  & x &  &  \\
2-10 & Stars change their color. & x &  & \\
2-11 & Massive stars can evolve into neutron stars. & x &  & \\
2-12 & Massive stars can evolve into black holes. & x &  & \\
2-13 & Stars fade and disappear over time.   &  & x & \cite{agan2004stellar}\\
2-14 & Supernova eruptions occur only in our galaxy.  & & x &  \\
2-15 & All stars end up in a supernova.  & & x &  \\
2-16 & A supernova immediately destroys a large part of the galaxy.  & & x &  \\
\bottomrule
\end{tabularx}
\end{table*}

\begin{table*}[hbt]
\subsection{Sub-scale on domain 3: Properties of
Stars} \label{instrument:domain3}
\vspace{-20pt}
\caption{\label{Tab:RatingScale3} Overview of the rating scale items related to the sub-scale on domain 3. The table includes the item numbers (No.), the statements themselves, indications of whether the statements align with the scientific view or oppose it, and any references or sources of inspiration.}
\begin{tabularx}{\textwidth}{p{0.05\textwidth}p{0.5\textwidth}p{0.15\textwidth}p{0.15\textwidth}p{0.1\textwidth}}
\toprule
No. & Item & In line with scientific view & Opposing scientific view & Reference / Inspired by \\
\midrule	
3-1 & All stars are the same size.  & & x &  \\
3-2 & There are stars that are larger than the Sun. & x &  & \cite{agan2004stellar} \\
3-3 & All stars have the same mass. &  & x & \\
3-4 & Stars underlie gravitational pull.   & x &  &\\
3-5 & Stars exert attraction on things.  & x &  & \\
3-6 & All stars have about the same distance from Earth as the Moon.& & x & \\
3-7 & The brightest stars are closest to Earth.  & & x & \cite{agan2004stellar} \\
3-8 & All stars are the same distance from Earth.  &  & x & \\
3-9 & Stars are farther away from Earth than the Sun.  & x &  &  \\
3-10 & The distance between stars is about the same as the distance between planets. & &x  & \\
3-11 & All stars are stationary – they are fixed in the sky and do not move.  &  & x & \cite{plummer2009cross,plummer2009early,kuccukozer2009effect} \\
3-12 & During the course of a night we see different stars.    & x  & & \cite{plummer2009cross}\\
3-13 & Stars seem to rise and set.   & x &  & \cite{comins2001heavenly,agan2004stellar}   \\
3-14 & The observable motion of the stars is a result of the Earth's rotation around its own axis.   & x & & \cite{lelliott2010big,anantasook2018thai}  \\
3-15 & Stars don’t rotate.   & & x & \cite{comins2001heavenly,agan2004stellar}  \\
\bottomrule
\end{tabularx}
\end{table*}

\begin{table*}[hbt]
\subsection{Sub-scale on domain 4: (Sub-)Stellar objects} \label{instrument:domain4}
\vspace{-20pt}
\caption{\label{Tab:RatingScale4} Overview of the rating scale items related to the sub-scale on domain 4. The table includes the item numbers (No.), the statements themselves, indications of whether the statements align with the scientific view or oppose it, and any references or sources of inspiration.}
\begin{tabularx}{\textwidth}{p{0.03\textwidth}p{0.5\textwidth}p{0.15\textwidth}p{0.15\textwidth}p{0.1\textwidth}}
\toprule
No. & Item & In line with scientific view & Opposing scientific view & Reference / Inspired by \\
\midrule	
4-1 & Two orbiting stars (binary star system) would quickly collide.  & & x & \\
4-2 & There are long-standing binary star systems. & x &  & \\
4-3 & Brown dwarfs describe the final stage of a star. &  & x & \\
4-4 & White dwarfs are planets.   &  & x &\\
4-5 & White dwarfs are suns.  & x &  & \\
4-6 & White dwarfs are stars. & x &  & \\
4-7 & All stars end up as white dwarfs.  & & x & \\
4-8 & Pulsars are stars that alternately emit light of different intensities. &  & x & \\
\bottomrule
\end{tabularx}
\end{table*}

\newpage

\begin{table*}[hbt]
\subsection{Sub-scale on domain 5: Spectral aspects} \label{instrument:domain5}
\vspace{-20pt}
\caption{\label{Tab:RatingScale5} Overview of the rating scale items related to the sub-scale on domain 5. The table includes the item numbers (No.), the statements themselves, indications of whether the statements align with the scientific view or oppose it, and any references or sources of inspiration.}
\begin{tabularx}{\textwidth}{p{0.03\textwidth}p{0.5\textwidth}p{0.15\textwidth}p{0.15\textwidth}p{0.1\textwidth}}
\toprule
No. & Item & In line with scientific view & Opposing scientific view & Reference / Inspired by \\
\midrule	
5-1 & All stars have the same color.  & & x & \\
5-2 & All stars are white. &  & x & \\
5-3 & All stars are yellow.  &  & x & \cite{comins2001heavenly,agan2004stellar}\\
5-4 & There are blue stars.   & x &  &\\
5-5 & Stars emit many colors of light.  & x &  & \\
5-6 & The brightness of stars is constant.&  & x  & \\
5-7 & It is said that stars twinkle because they change their brightness.   & & x & \cite{comins2001heavenly,agan2004stellar}\\
\bottomrule
\end{tabularx}
\end{table*}

\begin{table*}
\subsection{Distribution of students' responses on all items} 
\caption{\label{tab:d1} Share of responses to all items of domain 1 that are in line with scientific views ($+$), opposing scientific views ($-$) or abstained form voting ($\circ$), for all three cohorts.}
\begin{ruledtabular}

\begin{tabular}{ccccc|ccccc}
Item & Cohort & $-$ & $\circ$ & $+$ & Item & Cohort & $-$ & $\circ$ & $+$\tabularnewline
\hline 
\hline 
\multirow{3}{*}{1-1} & 1 & 41.9 & 10.1 & 48.0 & \multirow{3}{*}{1-6} & 1 & 46.6 & 32.4 & 20.9\tabularnewline
 & 2 & 13.9 & 2.0 & 84.1 &  & 2 & 40.4 & 25.8 & 33.8\tabularnewline
 & 3 & 9.0 & 0.0 & 91.0 &  & 3 & 29.9 & 28.4 & 41.8\tabularnewline 
\hline
\multirow{3}{*}{1-2} & 1 & 36.5 & 6.8 & 67.6 & \multirow{3}{*}{1-7} & 1 & 19.6 & 10.1 & 70.3\tabularnewline
& 2 & 12.6 & 0.7 & 86.6 &  & 2 & 11.9 & 6.0 & 82.1\tabularnewline
& 3 & 11.9 & 0.0 & 88.1 &  & 3 & 10.4 & 1.5 & 88.1\tabularnewline
\hline 
\multirow{3}{*}{1-3} & 1 & 25.7 & 6.8 & 67.6 & \multirow{3}{*}{1-8} & 1 & 10.8 & 6.8 & 82.4\tabularnewline
& 2 & 21.2 & 3.3 & 75.5 &  & 2 & 11.3 & 2.0 & 86.6\tabularnewline
& 3 & 9.0 & 0.0 & 91.0 &  & 3 & 3.0 & 0.0 & 97.0\tabularnewline
\hline 
\multirow{3}{*}{1-4} & 1 & 32.4 & 17.6 & 50.0 & \multirow{3}{*}{1-9} & 1 & 60.1 & 14.9 & 25.0\tabularnewline
& 2 & 46.1 & 10.6 & 43.0 &  & 2 & 37.7 & 6.0 & 56.3\tabularnewline
& 3 & 41.8 & 4.5 & 53.7 &  & 3 & 29.9 & 4.5 & 65.7\tabularnewline
\hline 
\multirow{3}{*}{1-5} & 1 & 75.0 & 10.1 & 14.9 & \multicolumn{1}{c}{} & \multicolumn{1}{c}{} & \multicolumn{1}{c}{} & \multicolumn{1}{c}{} & \multicolumn{1}{c}{}\tabularnewline
& 2 & 70.9 & 0.0 & 29.1 & \multicolumn{1}{c}{} & \multicolumn{1}{c}{} & \multicolumn{1}{c}{} & \multicolumn{1}{c}{} & \multicolumn{1}{c}{}\tabularnewline
 & 3 & 62.7 & 0.0 & 37.3 & \multicolumn{1}{c}{} & \multicolumn{1}{c}{} & \multicolumn{1}{c}{} & \multicolumn{1}{c}{} & \multicolumn{1}{c}{}\tabularnewline
\end{tabular}
\end{ruledtabular}
\end{table*}

\begin{table*}
\caption{\label{tab:d2} Share of responses to all items of domain 2 that are in line with scientific views ($+$), opposing scientific views ($-$) or abstained form voting ($\circ$), for all three cohorts.}
\begin{ruledtabular}

\begin{tabular}{ccccc|ccccc}
Item & Cohort & $-$ & $\circ$ & $+$ & Item & Cohort & $-$ & $\circ$ & $+$\tabularnewline
\hline
\hline
\multirow{3}{*}{2-1} & 1 & 52.0 & 18.2 & 29.7 & \multirow{3}{*}{2-9} & 1 & 14.2 & 20.9 & 64.9\tabularnewline
 & 2 & 37.7 & 18.5 & 43.7 &  & 2 & 24.5 & 13.2 & 62.3\tabularnewline
 & 3 & 23.9 & 11.9 & 64.2 &  & 3 & 11.9 & 9.0 & 79.1\tabularnewline
\hline 
\multirow{3}{*}{2-2} & 1 & 71.6 & 10.1 & 18.2 & \multirow{3}{*}{2-10} & 1 & 50.7 & 32.4 & 16.9\tabularnewline
 & 2 & 41.7 & 9.3 & 49.0 &  & 2 & 49.0 & 15.2 & 35.8\tabularnewline
 & 3 & 25.4 & 3.0 & 71.6 &  & 3 & 38.8 & 9.0 & 79.1\tabularnewline
\hline 
\multirow{3}{*}{2-3} & 1 & 50.0 & 18.2 & 31.8 & \multirow{3}{*}{2-11} & 1 & 7.4 & 50.7 & 41.9\tabularnewline
 & 2 & 24.5 & 15.2 & 60.3 &  & 2 & 12.6 & 56.3 & 31.1\tabularnewline
 & 3 & 11.9 & 9.0 & 79.1 &  & 3 & 10.4 & 44.8 & 44.8\tabularnewline
\hline 
\multirow{3}{*}{2-4} & 1 & 7.4 & 24.3 & 68.2 & \multirow{3}{*}{2-12} & 1 & 37.2 & 35.1 & 27.7\tabularnewline
 & 2 & 11.3 & 9.9 & 78.8 &  & 2 & 22.5 & 27.8 & 49.7\tabularnewline
 & 3 & 4.5 & 9.0 & 86.6 &  & 3 & 16.4 & 22.4 & 61.2\tabularnewline
\hline 
\multirow{3}{*}{2-5} & 1 & 33.1 & 27.0 & 39.9 & \multirow{3}{*}{2-13} & 1 & 63.5 & 10.8 & 25.7\tabularnewline
 & 2 & 29.1 & 24.5 & 46.4 &  & 2 & 58.3 & 11.3 & 30.5\tabularnewline
 & 3 & 13.4 & 20.9 & 65.7 &  & 3 & 53.7 & 10.4 & 35.8\tabularnewline
\hline 
\multirow{3}{*}{2-6} & 1 & 18.2 & 11.5 & 70.3 & \multirow{3}{*}{2-14} & 1 & 27.7 & 29.7 & 42.6\tabularnewline
 & 2 & 9.9 & 6.0 & 84.1 &  & 2 & 11.9 & 25.2 & 62.9\tabularnewline
 & 3 & 9.0 & 1.5 & 89.6 &  & 3 & 13.4 & 14.9 & 71.6\tabularnewline
\hline 
\multirow{3}{*}{2-7} & 1 & 14.9 & 20.9 & 64.2 & \multirow{3}{*}{2-15} & 1 & 38.5 & 25.0 & 36.5\tabularnewline
 & 2 & 19.2 & 5.3 & 75.5 &  & 2 & 38.4 & 30.5 & 31.1\tabularnewline
 & 3 & 16.4 & 6.0 & 77.6 &  & 3 & 34.3 & 17.9 & 47.8\tabularnewline
\hline 
\multirow{3}{*}{2-8} & 1 & 3.4 & 17.6 & 79.1 & \multirow{3}{*}{2-16} & 1 & 50.0 & 21.6 & 28.4\tabularnewline
 & 2 & 5.3 & 2.6 & 92.1 &  & 2 & 35.1 & 36.4 & 28.5\tabularnewline
 & 3 & 6.0 & 4.5 & 89.6 &  & 3 & 47.8 & 22.4 & 29.9\tabularnewline
\end{tabular}
\end{ruledtabular}
\end{table*}

\begin{table*}
\caption{\label{tab:d3} Share of responses to all items of domain 3 that are in line with scientific views ($+$), opposing scientific views ($-$) or abstained form voting ($\circ$), for all three cohorts.}
\begin{ruledtabular}

\begin{tabular}{ccccc|ccccc}
Item & Cohort & $-$ & $\circ$ & $+$ & Item & Cohort & $-$ & $\circ$ & $+$\tabularnewline
\hline 
\hline 
\multirow{3}{*}{3-1} & 1 & 0.0 & 14.2 & 85.8 & \multirow{3}{*}{3-9} & 1 & 35.1 & 20.9 & 43.9\tabularnewline
 & 2 & 2.0 & 2.0 & 96.0 &  & 2 & 33.1 & 15.2 & 51.7\tabularnewline
 & 3 & 1.5 & 0.0 & 98.5 &  & 3 & 35.8 & 10.4 & 53.7\tabularnewline
\hline 
\multirow{3}{*}{3-2} & 1 & 48.6 & 14.2 & 37.2 & \multirow{3}{*}{3-10} & 1 & 17.6 & 32.4 & 50.0\tabularnewline
 & 2 & 28.5 & 7.3 & 64.2 &  & 2 & 13.9 & 23.2 & 62.9\tabularnewline
 & 3 & 10.4 & 6.0 & 83.6 &  & 3 & 7.5 & 17.9 & 74.6\tabularnewline
\hline 
\multirow{3}{*}{3-3} & 1 & 6.8 & 25.0 & 68.2 & \multirow{3}{*}{3-11} & 1 & 18.2 & 10.1 & 71.6\tabularnewline
 & 2 & 6.6 & 4.6 & 88.7 &  & 2 & 26.5 & 11.3 & 62.3\tabularnewline
 & 3 & 4.5 & 0.0 & 95.5 &  & 3 & 14.9 & 9.0 & 76.1\tabularnewline
\hline 
\multirow{3}{*}{3-4} & 1 & 43.9 & 18.9 & 37.2 & \multirow{3}{*}{3-12} & 1 & 25.0 & 14.2 & 60.8\tabularnewline
 & 2 & 29.1 & 22.5 & 48.3 &  & 2 & 15.9 & 8.6 & 75.5\tabularnewline
 & 3 & 37.3 & 13.4 & 49.3 &  & 3 & 11.9 & 6.0 & 82.1\tabularnewline
\hline 
\multirow{3}{*}{3-5} & 1 & 25.0 & 14.9 & 60.1 & \multirow{3}{*}{3-13} & 1 & 35.8 & 20.9 & 43.2\tabularnewline
 & 2 & 25.8 & 25.8 & 48.3 &  & 2 & 43.0 & 20.5 & 36.4\tabularnewline
 & 3 & 19.4 & 11.9 & 68.7 &  & 3 & 34.3 & 11.9 & 53.7\tabularnewline
\hline 
\multirow{3}{*}{3-6} & 1 & 14.2 & 21.6 & 64.2 & \multirow{3}{*}{3-14} & 1 & 25.7 & 28.4 & 43.2\tabularnewline
 & 2 & 9.3 & 7.3 & 83.4 &  & 2 & 18.5 & 21.9 & 59.6\tabularnewline
 & 3 & 1.5 & 3.0 & 95.5 &  & 3 & 11.9 & 14.9 & 73.1\tabularnewline
\hline 
\multirow{3}{*}{3-7} & 1 & 39.9 & 10.8 & 49.3 & \multirow{3}{*}{3-15} & 1 & 43.9 & 18.9 & 37.2\tabularnewline
 & 2 & 42.4 & 8.6 & 49.0 &  & 2 & 33.1 & 22.5 & 44.4\tabularnewline
 & 3 & 28.4 & 4.5 & 67.2 &  & 3 & 26.9 & 11.9 & 61.2\tabularnewline
\hline 
\multirow{3}{*}{3-8} & 1 & 0.0 & 22.3 & 77.7 & \multicolumn{1}{c}{\multirow{3}{*}{}} & \multicolumn{1}{c}{} & \multicolumn{1}{c}{} & \multicolumn{1}{c}{} & \multicolumn{1}{c}{}\tabularnewline
 & 2 & 6.0 & 2.0 & 92.1 &  & \multicolumn{1}{c}{} & \multicolumn{1}{c}{} & \multicolumn{1}{c}{} & \multicolumn{1}{c}{}\tabularnewline
 & 3 & 1.5 & 1.5 & 97.0 &  & \multicolumn{1}{c}{} & \multicolumn{1}{c}{} & \multicolumn{1}{c}{} & \multicolumn{1}{c}{}\tabularnewline
\end{tabular}
\end{ruledtabular}
\end{table*}

\begin{table*}
\centering
\caption{\label{tab:d4} Share of responses to all items of domain 4 that are in line with scientific views ($+$), opposing scientific views ($-$) or abstained form voting ($\circ$), for all three cohorts.}
\begin{ruledtabular}
\begin{tabular}{ccccc|ccccc}
Item & Cohort & $-$ & $\circ$ & $+$ & Item & Cohort & $-$ & $\circ$ & $+$\tabularnewline
\hline 
\hline 
\multirow{3}{*}{4-1} & 1 & 29.7 & 31.8 & 38.5 & \multirow{3}{*}{4-5} & 1 & 38.5 & 48.0 & 13.5\tabularnewline
 & 2 & 30.5 & 37.1 & 32.5 &  & 2 & 28.5 & 45.0 & 26.5\tabularnewline
 & 3 & 25.4 & 29.9 & 44.8 &  & 3 & 47.8 & 26.9 & 25.4\tabularnewline
\hline 
\multirow{3}{*}{4-2} & 1 & 21.6 & 21.6 & 56.8 & \multirow{3}{*}{4-6} & 1 & 17.6 & 43.9 & 38.5\tabularnewline
 & 2 & 13.9 & 43.0 & 43.0 &  & 2 & 16.6 & 44.4 & 39.1\tabularnewline
 & 3 & 13.4 & 29.9 & 56.7 &  & 3 & 9.0 & 20.9 & 70.1\tabularnewline
\hline 
\multirow{3}{*}{4-3} & 1 & 39.2 & 36.5 & 24.3 & \multirow{3}{*}{4-7} & 1 & 20.9 & 47.3 & 31.8\tabularnewline
 & 2 & 30.5 & 48.3 & 21.2 &  & 2 & 20.5 & 48.3 & 31.1\tabularnewline
 & 3 & 37.3 & 34.3 & 28.4 &  & 3 & 26.9 & 32.8 & 40.3\tabularnewline
\hline 
\multirow{3}{*}{4-4} & 1 & 10.8 & 51.4 & 37.8 & \multirow{3}{*}{4-8} & 1 & 36.5 & 41.9 & 21.6\tabularnewline
 & 2 & 5.3 & 45.7 & 49.0 &  & 2 & 36.5 & 55.0 & 11.9\tabularnewline
 & 3 & 10.4 & 26.9 & 25.4 &  & 3 & 41.8 & 44.8 & 13.4\tabularnewline
\end{tabular}
\end{ruledtabular}
\end{table*}

\begin{table*}
\centering
\caption{\label{tab:d5} Share of responses to all items of domain 5 that are in line with scientific views ($+$), opposing scientific views ($-$) or abstained form voting ($\circ$), for all three cohorts.}
\begin{ruledtabular}
\begin{tabular}{ccccc|ccccc}
Item & Cohort & $-$ & $\circ$ & $+$ & Item & Cohort & $-$ & $\circ$ & $+$\tabularnewline
\hline 
\hline 
\multirow{3}{*}{5-1} & 1 & 20.3 & 10.8 & 68.9 & \multirow{3}{*}{5-5} & 1 & 15.5 & 14.4 & 69.6\tabularnewline
 & 2 & 13.9 & 7.9 & 78.1 &  & 2 & 18.5 & 12.6 & 68.9\tabularnewline
 & 3 & 4.5 & 7.5 & 88.1 &  & 3 & 11.9 & 7.5 & 80.6\tabularnewline
\hline 
\multirow{3}{*}{5-2} & 1 & 14.2 & 10.8 & 75.0 & \multirow{3}{*}{5-6} & 1 & 38.5 & 22.3 & 39.2\tabularnewline
 & 2 & 13.9 & 8.6 & 77.5 &  & 2 & 19.2 & 6.6 & 74.2\tabularnewline
 & 3 & 9.0 & 7.5 & 83.6 &  & 3 & 9.0 & 4.5 & 86.6\tabularnewline
\hline 
\multirow{3}{*}{5-3} & 1 & 10.8 & 18.2 & 70.9 & \multirow{3}{*}{5-7} & 1 & 45.9 & 29.1 & 25.0\tabularnewline
 & 2 & 13.2 & 9.9 & 76.8 &  & 2 & 39.1 & 19.2 & 41.7\tabularnewline
 & 3 & 4.5 & 7.5 & 88.1 &  & 3 & 25.4 & 19.4 & 55.2\tabularnewline
\hline 
\multirow{3}{*}{5-4} & 1 & 14.9 & 21.6 & 63.5 & \multicolumn{1}{c}{\multirow{3}{*}{}} & \multicolumn{1}{c}{} & \multicolumn{1}{c}{} & \multicolumn{1}{c}{} & \multicolumn{1}{c}{}\tabularnewline
 & 2 & 30.5 & 15.9 & 53.6 &  & \multicolumn{1}{c}{} & \multicolumn{1}{c}{} & \multicolumn{1}{c}{} & \multicolumn{1}{c}{}\tabularnewline
 & 3 & 19.4 & 10.4 & 70.1 &  & \multicolumn{1}{c}{} & \multicolumn{1}{c}{} & \multicolumn{1}{c}{} & \multicolumn{1}{c}{}\tabularnewline
\end{tabular}
\end{ruledtabular}
\end{table*}

\end{document}